\documentclass[twocolumn,showpacs,preprintnumbers,amsmath,amssymb]{revtex4}
\usepackage{graphicx}% Include figure files
\usepackage{dcolumn}% Align table columns on decimal point
\usepackage{bm}% bold math

\begin{document}

\preprint{}

\title{The Conventionality of Synchronization and\\the Causal Structure of Quantum Mechanics}

\author{James Carrubba}
 \email{carrubba@uiuc.edu}
\affiliation{%
University Laboratory High School\\
University of Illinois\\
1212 West Springfield\\
Urbana IL 61821
}%

\date{\today}

\begin{abstract}
Measuring velocities requires the synchronization
of spatially-separated clocks.  Because this synchronization
must precede the determination of velocities, no
system of clock synchronization---such as that based
on Einstein's presumption of light-speed isotropy---can
ever be founded on an experimentally-validated velocity.
I argue that this very old observation, which lingers in the
philosophical literature under the heading ``Conventionality of
Synchronization,'' suggests an explanation of why ``spooky''
quantum correlations can transfer no information at any
speed, superluminal or otherwise.  This work constitutes
the first application of the Conventionality doctrine outside
of Relativity itself.
\end{abstract}

% PACS, the Physics and Astronomy Classification Scheme.
\maketitle

Faster-than-light correlation of measurements, implicit
in accepted quantum mechanics\cite{epr,Bell:1964kc,PhysRevLett.23.880}, is by now
supported experimentally\cite{asp,kwiat,tittel}.  Because no information
is transmitted superluminally, these correlations technically
do not violate the proscriptions of Special Relativity.
Nevertheless, the idea of {\em any\/} superluminal effect
seems offensive to the spirit of Relativity; these two
fundamental theories appear therefore to coexist uneasily.

Standards of length and time must underlie any quantitative
description of nature.  If it is true that Relativity provides
the correct description of these standards, then rather
than looking to ``reconcile'' Relativity and Quantum Mechanics,
we should prefer that Relativity {\em compel\/} the
space-time structure of Quantum Mechanics.

It does not appear that the Special Theory offers any insights
into the structure of Quantum Mechanics.  I suggest, however,
that a class of general coordinate transformations, applied
in flat space-time, explains why information can {\em never\/}
be transmitted at any speed---neither superluminal nor
subluminal---by quantum measurement.

\section{\label{sec:clocksynch}Clock Synchronization}

The observation most relevant to our
work---that it is circular to ``verify'' the one-way
speed of light using clocks synchronized on the assumption of
light-speed isotropy---dates back many decades\cite{reich,grun},
and is sometimes abbreviated the ``Conventionality of Synchronization.''
Winnie\cite{win1,win2} derived a limited set of Lorentz-like
transforms relating frames in which clocks are synchronized
by some procedure other than Einstein's.  More useful, for
our purposes, is the fact that resynchronizations can
be effected by similarity transforms, as in
\begin{equation} \label{eq:alpha}
\begin{pmatrix} dt' \\ dx' \end{pmatrix}
 = \begin{pmatrix} 1 & \alpha \\ 0 & 1 \end{pmatrix}
\begin{pmatrix} dt \\ dx \end{pmatrix}
\end{equation}
relating an Einstein-synchronized unprimed frame to a primed
frame in which, it is easy to see, light has velocity
\begin{equation}
c' = \frac{\pm c}{1 \pm \alpha c}
\end{equation}
in the positive (top sign) and negative directions.  Here
and henceforth,
$c\approx 3\times 10^8$ meters per second denotes the
well-known speed which, I argue, is necessarily measured
only for round-trip trajectories.

About this three points should be made clear.  First, the
Conventionality doctrine---the claim that light's
round-trip speed, measured countless times in labs
innumerable, offers no information about light's
one-way speed---describes nothing peculiar to light.
The freedom to resynchronize afflicts all one-way
velocities.  The child who
runs to the east at 2 meters per second, touches base,
and runs back to the west at two meters per second,
could also be described as running at 4 meters per
second to the east and 4/3 meters per second to the
west.  (A trivial point regarding Relativity:
a stationary observer, watching the child
pass meter markings and checking against a stopwatch,
cannot assert symmetry of velocities, because
this observer really isn't measuring when the child
reaches meter markings!\footnote{The stationary
observer is of course measuring when light
reaches her eye.  Her measurements,
being read off a single clock, are not subject
to synchronization ambiguities.  It is easy to see
that a resynchronization
transforming the child's velocities $v_1=-v_2=2$
into $v'_1=4$ and $v'_2=-4/3$ would affect the
east and west speeds of light correspondingly
and thus guarantee the invariance of her observations.})
The issue here is the
very definition of {\em velocity.}  No one-way velocity
of anything can ever be measured until clocks at different
points are synchronized.  Round-trip speeds are, of
course, indifferent to synchronizations, as is too
the elapsed time between arrivals of
different objects at a single clock.%\footnote{The
%anisotropy addressed here, being an experimentally
%inaccessible consequence of the very definition of velocity,
%is totally different from that anisotropy possibly
%associated with violations of Local Lorentz Invariance,
%as for instance in SEE PRL 05.}

Second:  it may sound reasonable to synchronize clocks at
rest in one frame by using a third clock, carried from
one to the other, and invoking the Lorentz transformations
to account for time-dilation.  An obvious problem is that,
until clocks have been synchronized, we don't know what
velocity $v$ to plug in to the Lorentz transformations.
Further, the Lorentz transformations presume
the Einstein synchronization; invoking them would
``smuggle'' isotropy into the mathematics.\footnote{There
are schemes to synchronize separated clocks by taking the speed
$v\rightarrow 0$ for the synchronizing clock---that is,
to avoid time dilation effects.  These schemes miss the
point that we can only assert \emph{by definition} the
effect of motion on synchrony.}

Lastly:  could there be some physical result, possibly
from some other branch of physics, which would compel
Einstein's symmetric synchronization?  I argue
that this is unlikely.  No physical result which
depends on measurements of time at different points
can precede the synchronization of clocks.  All
claims must be checked, again, that they not sneak
conventional synchronizations into putatively
non-conventional results.

Does Conventionality harbor philosophical idealism?  It may look
(for instance) as if human opinion intrudes
on reality enough to make light travel
to the right as quickly or slowly as we please, with this light
obligingly making up the difference during a return trip
to left, always conspiring to realize a total travel time of $2L/c$!

Quite to the contrary, this work strikes a perfectly
realist pose.  Light will do whatever light is going
to do (which evidently includes making round trip
journeys in time $t=2L/c$); like the word ``velocity''
itself, terms like ``faster'' and ``slower'' have no
precise meaning in this context outside of an agreed-upon
synchronization.  It makes sense to say that a photon
reached $x=L$ faster than a proton also emitted from the
space-time origin; but it makes no sense---is simply
not defined---to say that the photon travels ``faster''
if the clock at $x=L$ is synchronized so that the photon's
arrival is measured to be at time $t=0.6 L/c$, than if
it travels rightward in an Einstein-synchronized frame.

Further discussion should be sought in the
literature\cite{anderson-vetharaniam-stedman,sep-spacetime-convensimul}.

\section{Information}

Relativity forbids the superluminal transmission of
information.  But what counts as information?  For
our purposes, it is convenient to observe that
the transmission of a signal forces a causal ordering of events.
If a signal can be sent from event ${\bf e_1}$ to event
${\bf e_2}$, then the first event could have caused
the second, regardless of how local clocks
are synchronized.  If these
events are measurements---say, observer {\bf A}
applying operator ${\cal O}_A$ to a left-moving
particle, and observer {\bf B} applying operator ${\cal O}_B$
to a particle moving rightward---then the commutativity
$[{\cal O}_A,{\cal O}_B]=0$
of these measurements precludes the transmission of
information between observers {\bf A} and {\bf B}.
That is:  if it makes no physical difference which
measurement is made first, then neither event could
have caused the other.  We might well wonder under
what circumstances $[{\cal O}_A,{\cal O}_B]=0$.

Consider a situation where the left- and right-moving
particles do not interact.  The time-development of the
quantum state is governed by a Hamiltonian $H = H_A + H_B$,
where $H_A$ and $H_B$
commute with each other (and therefore with $H,$ too).
%was previously a paragraph break
Measurements ${\cal O}_A$ and ${\cal O}_B$ will be made
on the left and right particles, respectively.  We make the
additional assumption that $[H_A,{\cal O}_B] = [H_B,{\cal O}_A] = 0$:
a measurement on one particle will not affect the evolution of the
other.  These three assumptions of commutativity hold for
EPR experiments.

We calculate the transition amplitude ${\cal M}$ connecting
an incoming state $|\psi_{\mathit in}>$ at time $t_{\mathit in}$
and an outgoing state $|\psi_{\mathit out}>$ at $t_{\mathit out}$
through the measurements ${\cal O}_A$ and ${\cal O}_B$.
For measurement times $t_A < t_B$ we write, according to
the usual prescriptions of quantum mechanics:
\begin{widetext}
\begin{eqnarray}
%<\psi_{\mathit out}|\psi_{\mathit in}> &=&
{\cal M} &=&
<\psi_{\mathit out}|e^{-i H (t_{\mathit out} - t_B)} {\cal O}_B
e^{-i H (t_B - t_A)} {\cal O}_A
e^{-i H (t_A - t_{\mathit in})}
|\psi_{\mathit in}> \nonumber\\
&=&
<\psi_{\mathit out}|e^{-i H_B (t_{\mathit out} - t_B)} {\cal O}_B
e^{-i H_B (t_B - t_{\mathit in})} %\nonumber \\ & & \times \,\,\,\,
e^{-i H_A (t_{\mathit out} - t_A)}
{\cal O}_A e^{-i H_A (t_A - t_{\mathit in})}|\psi_{\mathit in}> \nonumber \\
&=&
<\psi_{\mathit out}| {\cal O}_B(t_B) {\cal O}_A(t_A)
|\psi_{\mathit in}> \label{Afirst}
\end{eqnarray}
\end{widetext}
Along the way to equation (\ref{Afirst}) we have made use of the
commutators involving $H_A$ and $H_B$.

If the theory is to remain covariant under the transformations
(\ref{eq:alpha}), then we must be allowed to
resynchronize clocks arbitrarily, even to the extent that $t_A' > t_B'$
for {\em time-like separated measurements.}  A sequence of steps
analogous to those leading to (\ref{Afirst}) shows
\begin{equation}
{\cal M'} =
<\psi_{\mathit out}| {\cal O}_A(t_A') {\cal O}_B(t_B')
|\psi_{\mathit in}> \label{Bfirst}
\end{equation}
In the limits $t_A\rightarrow t_B$ and $\alpha\rightarrow 0$,
expression (\ref{Afirst}) must approach (\ref{Bfirst}); thus $[{\cal O}_A,{\cal O}_B]=0$.
(Since there are only finitely many degrees of freedom here,
we do not expect to encounter singularities in taking the limits.)

Even readers comfortable with the Conventionality doctrine will find
at least one part of this discussion objectionable.  It is typical
in the literature\footnote{Winnie\cite{win1,win2} limits his parameter
$0<\epsilon<1$, where $\epsilon=1/2$ corresponds to the Einstein convention.}
to limit $|\alpha c|<1$ in equation
(\ref{eq:alpha}).  This preserves the
``common-sense'' restriction that the departure time
$t_A$ of a signal leaving ${\bf A}$ and arrival time
$t_B$ at ${\bf B}$ always obey $t_A < t_B$, regardless
of how clocks at those locations are otherwise synchronized.
Any pair of clocks---according to this reasoning---should reflect causality.
It might sound like a cautious compromise to accept
the arguments leading to eq.~(\ref{eq:alpha})
for space-like separated events but reject the
``radical'' (arbitrary-$\alpha$) Conventionality espoused above.

On general grounds, I reply that tolerating the
{\em widest possible range of mathematical descriptions\/} has
always proven to be the wisest approach to physics,
not just for the sake of convenience---{\em e.g.,}
the center of mass frame may make a cross section easier to calculate---but
because the {\em very freedom\/} to choose different descriptions
may itself be important (for instance, building particle
representations out of the Poincar{\'e} group).

\begin{figure}
\includegraphics{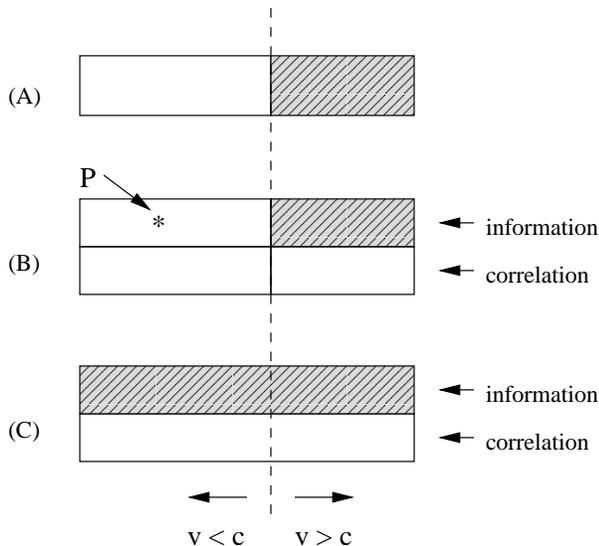}
\caption{\label{onlyfigure}  In this figure, in which
the (normal) Einstein-synchronziation is presumed, the horizontal
axis indicates velocity; $v=c$ along the dashed vertical line.
Shading indicates physically prohibited regions.
(A) In the most straightforward interpretation of Relativity,
superluminal {\em (shaded region)} effects are prohibited.  Causal
signals would plot along or to the left of the dashed line.
(B) Quantum entanglement forces a distinction between the superluminal
passage of information, which is still forbidden, and ``spooky'' superluminal {\em
correlation.}  Point $P$ exemplifies the {\em subluminal\/} passage of
information by quantum measurement; in standard interpretations, $P$ is
accepted as consistent with Relativity.
(C) This depicts the actual status of non-relativistic
quantum mechanics:  all transmission
of information by measurement is prohibited, but correlation is
allowed.  In the text, we argue that this picture---including the apparent
indifference to the speed of light---is compelled by Relativity.}
\end{figure}
%\end{minipage}

%\vspace*{\stretch{1}}

%\newpage

I argue that this is the
case here.  Sketch (A) of Fig.~\ref{onlyfigure} suggests the most naive
view of causality required by Relativity.  Superluminal
effects are proscribed (shaded region)\footnote{It is
a simple demonstration---accessible, in my experience,
to high school students in their second year of physics---that
a signal transmitted and received simultaneously in one
Einstein-synchronized frame can lead to information
traveling backward in time in another.
General coordinate transforms would perhaps be too ambitious.}.
Quantum mechanics compels a
more nuanced view, reflected in the distinction
(sketch (B) of Fig.~\ref{onlyfigure}) between
the superluminal passage of information (still prohibited)
and discomforting but apparently real superluminal correlations.

The problem with constraining $|\alpha c|<1$ is that quantum mechanics
is formally indifferent to the speed of light:  quantum
measurement refuses to send information at any speed,
superluminal or otherwise.  Allowing arbitrary
synchronizations---including those in which clocks
do not reflect causality---explains why points like
$P$ in (B) are prohibited, even though
seemingly acceptable according to the most naive requirements
of causality.  Accepting broadest synchronization arguments
leads to the physically correct picture, part (C) of
Fig.~\ref{onlyfigure}.

None of this discredits Einstein's original assumption
that the one-way speed of light equals the physical
constant $c$; it merely reinforces, I hope, the ontological
status of that assumption (see Section \ref{sec:clocksynch}).

\section{Interactions}
The manipulations leading to eq.~(\ref{Afirst})
are obviously invalid if there is
any interaction between the two particles,
as then $[H_A,H_B]\ne 0$.
Under such circumstances one can, of course, transmit
information.  But what
if we try to repeat these manipulations anyway?  That is,
may we invoke my claimed right to synchronize clocks
arbitrarily, so that the time order of the operators ${\cal O}_A$ and
${\cal O}_B$ becomes arbitrary, even in a situation where a
signal is undeniably sent from {\bf A} to {\bf B}?

That we may do this, even in the presence of interactions,
is straightforward to see within a manifestly covariant formalism.

Under (\ref{eq:alpha}),
$t^{\prime} = t + \bf{\alpha}\cdot\bf{x}, \bf{x^{\prime}}=\bf{x}$,
leading us to rewrite the metric
\begin{eqnarray*}
g & = &
\begin{pmatrix}
1 & 0 & 0 & 0\\ 0 & -1 & 0 & 0\\0 & 0 & -1 & 0\\0 & 0 & 0 & -1
\end{pmatrix}
\rightarrow
\\
g' & = &
\begin{pmatrix}
1 & -\alpha_1 &-\alpha_2 &-\alpha_3 \\
-\alpha_1 & \alpha_1^2-1 & \alpha_1 \alpha_2 & \alpha_1 \alpha_3 \\
-\alpha_2 & \alpha_1 \alpha_2 & \alpha_2^2-1 & \alpha_2 \alpha_3 \\
-\alpha_3 & \alpha_1 \alpha_3 & \alpha_2 \alpha_3 & \alpha_3^2-1
\end{pmatrix}
\end{eqnarray*}
which results in the product
\begin{equation*}
g_{\mu\nu}' \, dx^{\prime \mu} \, dx^{\prime \nu} = (dt' - \bf{\alpha}\cdot d\bf{x'})^2 -
d\bf{x}^{\prime 2}.
\end{equation*}
Since, by definition, proper times are unaffected by resynchronization
of spatially-separated clocks, one finds once again speeds of light equal to
\begin{equation*}
%c' = \frac{c}{|1 \pm \bf{c}\cdot\bf{\alpha}|}
c' = \frac{c}{1 \pm \bf{c}\cdot\bf{\alpha}}
\end{equation*}
which preserve the round trip light speed $c$, as they must.
But these resynchronizations act differently on $k_{\mu} = (\omega,\bf{k})$:
frequency (measured by a clock at a point) does not change under
a resynchronization, but wavelength generally does
(wavelength being defined as the distance between the {\em simultaneous\/}
locations of two adjacent peaks).
These considerations lead to
$\bf{k}' = \bf{k} + \mathbf{\alpha} \omega$,
$\omega' = \omega$, and from there to the four-vector dot product
$k \cdot k = \omega^{2} - \bf{k}^2
= \omega^{\prime 2} - (\bf{k}' - \bf{\alpha} \omega')^2$.
The product $k \cdot x = \omega t - \bf{k}\cdot\bf{x}$ equals, by simple substitution,
$k' \cdot x' = \omega' t' - \bf{k}'\cdot\bf{x}'$.

In covariant theories, the propagator
\begin{equation*}
 = \int d^4 k
 \frac{e^{-i k\cdot x}}{\omega^{2} - k^2 - m^2 + i \epsilon}
 \end{equation*}
 governs the causal structure.  According to the previous
 paragraph, it can be written as
 \begin{equation*}
  = \int d^4 k'
  \frac{e^{-i k'\cdot x'}}{\omega^{\prime 2} - (k' - \alpha\, \omega')^2 - m^2 + i
  \epsilon}
  \end{equation*}
  By change of integration variables to $k''=k'-\alpha\, \omega'$,
  $\omega''=\omega'$, we end up with the expression
  \begin{equation*}
   = \int d^4 k''
   \frac{e^{-i (\omega'' (t'-\bf{\alpha\,\cdot x'})- \bf{k''\cdot x'})}}
   {\omega^{\prime\prime 2} - k^{\prime\prime 2} - m^2 + i \epsilon}
   \end{equation*}
   If, in other words, we start with a propagator in some
   alternative ($\alpha \ne 0$) synchronization,
   standard analysis\cite{mandlshaw} leads trivially to a causal structure
   equivalent to the expected one (since $t=t'-\bf{\alpha \,\cdot x'}$).
That is to say:  in a covariant theory, the ``causally earlier''
operators are automatically pushed to the right, regardless
of how clocks along the way may be synchronized.

\section{\label{sec:Discussion}Discussion}
To be concrete, imagine a photon is emitted
from the space-time origin $(t,x)=(0,0)$.
It is detected at $x=L>0$.  In a frame of
reference synchronized according to Einstein's
convention, the detection takes place at $t=L/c$
as measured by a local clock.

\subsection{\label{sec:collapse}``When'' do wavefunctions collapse?}

Superluminal signals are aphysical, but in a sense
they are at least comprehendible.  Having settled on
(for instance) the Einstein synchronization, we understand
that a photon emitted from the origin is detected at
$(t=L/c,x=L)$.  A signal emitted from the space-time
origin arriving at $L$ before time $t=L/c$ violates causality,
but at least it can be described mathematically.

On the other hand, there appears to be no way at all
to quantify when a wavefunction collapses.  Much
experimental work has been dedicated
to establishing Bell-violating correlations
at space like-separated events.  In such cases, one
observer might argue that $A$ was measured before $B$,
and another observer might disagree; but as the formalism
is symmetric, it didn't matter who ``caused'' the
first collapse.  Perhaps more noteworthy is that no signal
is ever sent by quantum measurement, even when
causally acceptable.  There may be no doubt that
observer {\bf A} carried out her measurement before
{\bf B}---in the sense that {\bf B} received {\bf A}'s letter urging
him to finish---yet as far as non-relativistic Quantum
Mechanics is concerned, there is \emph{still} no
reason to assert that she (and not he) first caused the
collapse of the wavefunction!

One might well reflect on whether experiment is
better understood from the Conventionalist 
perspective or from within the Special Theory, which
by definition imposes Einstein's symmetric synchronization.
For instance, the very clever experiment
(see \cite{PhysRevA.63.022111} and references therein)
which did much to inspire the current work has at its
very core assumptions\footnote{From\cite{PhysRevA.63.022111}:
``Each reference frame determines a time ordering.  Hence,
in each reference frame one measurement takes place
before the other and can be considered as the trigger
(the cause) of the collapse.''  This is, of course, the
ultimate target of Conventionalist criticism.}
which, I believe, cannot be treated as physically meaningful.
Whatever one's opinion of the Conventionalist doctrine,
the results in \cite{PhysRevA.63.022111} seem to
diminish hope that Special Relativity's presumptions
regarding simultaneity offer any insight into
wavefunction collapse.

%I invite readers to attempt proof that the
%collapse didn't occur before \emph{both} measurements.

\subsection{\label{sec:covariance}Formalism of Covariance}
Because wave four-vectors $(\omega,\mathbf{k})$
and displacement four-vectors $(t,\mathbf{x})$
transform differently under resynchronizations, care
must be taken in the treatment of Lorentz indices.
Given how casually indices get raised and lowered
in the Special Theory, one wonders in retrospect whether
the formalism of covariance should not have been
regarded as suspiciously under-exploited.

\subsection{\label{sec:naturalclocks}Clock resynchronization?}

That quantum measurement does not violate causality
might come to be accepted in an \emph{ad hoc} way, by which
various quantum effects---e.g., the No-Clone Theorem, the
limits on information extractable by state-disturbing
measurements---rise up,
\emph{Deus ex machina}, to prevent acausal signaling.
The current work claims that the causality demanded
by Relativity follows systematically
from the application of general coordinate transforms.

There is, however, a deeper issue here.  Readers familiar
with the philosophical literature on Relativity may
be aware of the distinction between ``natural'' and
``coordinate'' clocks.  Clocks in accelerating frames
generally desynchronize; ``coordinate'' clocks are by
definition resynchronized at each infinitesimal boost,
so that at any instant between boosts they reflect
the Einstein-synchronization presumed by the Special
Theory.  In running freely, real or ``natural'' clocks
flout physicists' conventions.
In that sense, the current work argues for the utility
of founding physical theories on the readings of
real clocks, however they may be synchronized.

\bibliography{doebib}% Produces the bibliography via BibTeX.
\end{document}